# ARTICLE

# Surface Acoustic Wave Hemolysis Assay for Evaluating Stored Red Blood Cells

*Meiou Song[a,#], Colin C. Anderson[b,#], Nakul Sridhar[a], Julie A. Reisz[b], Leyla Akh[c], Yu Gao[a], Angelo D'Alessandro[b,d]\*, Xiaoyun Ding[a,c,e]\**

Blood transfusion remains a cornerstone of modern medicine, saving countless lives daily. Yet the quality of transfused blood varies dramatically among donors—a critical factor often overlooked in clinical practice. Rapid, benchtop, and cost-effective methods for evaluating stored red blood cells (RBCs) at the site of transfusion are lacking, with concerns persisting about the association between metabolic signatures of stored RBC quality and transfusion outcomes. Recent studies utilizing metabolomics approaches to evaluate stored erythrocytes find that donor biology (e.g., genetics, age, lifestyle factors) underlies the heterogeneity associated with blood storage and transfusion. The appreciation of donor-intrinsic factors provides opportunities for precision transfusion medicine approaches for the evaluation of storage quality and prediction of transfusion efficacy. Here we propose a new platform, the Surface Acoustic Wave Hemolysis Assay (SAW-HA), for on-site evaluation of stored RBCs utilizing SAW Hemolysis Temperature (SAWHT) as a marker for RBC quality. We report SAWHT as a mechanism-dependent reproducible methodology for evaluating stored human RBCs up to 42 days. Our results define unique signatures for SAW hemolysis and metabolic profiles in RBCs from two of the six donors in which high body mass index (BMI) and RBC triglycerides associated with increased susceptibility to hemolysis. Metabolic age of the stored RBCs – a recently appreciated predictor of post-transfusion efficacy – reveal that RBCs from the two low SAWHT units were characterized by disrupted redox control, deficient tryptophan metabolism, and high BMI. Together, these findings indicate the potential of the SAW-HA as a point-of-care analysis for transfusion medicine.

## Introduction

Blood transfusion is a life-saving medical intervention for millions of recipients worldwide. Packed red blood cells (pRBCs) – the most widely transfused blood product – are stored for up to 42 days under refrigeration in additive solutions. However, stored RBCs undergo progressive metabolic and structural changes, collectively termed storage lesions.[1] These lesions, driven by oxidative stress and metabolic alterations, ultimately result in membrane damage, reduced deformability,[2] and hemolysis, along with an impaired capacity to deliver oxygen,[3] contributing to inflammatory sequelae and vascular dysfunction.[4] Notably, RBC deterioration does not occur uniformly; donor-specific factors such as sex, age, body mass index (BMI), genetic background, and metabolism significantly influence how RBCs respond to storage.[5–9] Current transfusion practices rely on the chronological storage age of RBCs, a metric that fails to capture metabolic age[10] — a function of RBC biophysical and biochemical integrity — which varies widely among donors. Without a benchtop assay that is easy for medical personnel to learn, use, and interpret quickly, transfusions may be non-optimized, potentially leading to poor outcomes in vulnerable patients.

Recent advances in -omics technologies, especially metabolomics, have provided new insights into RBC storage quality.[7,11–13] These approaches have identified potential biomarkers for transfusion effectiveness by characterizing metabolic changes in stored RBCs. For instance, biomarkers of oxidative stress (glutathione (GSH) and its synthetic intermediates) correlate strongly with oxidative hemolysis within stored RBCs.[14–16] Hypoxanthine, a marker of oxidant stress, accumulates nonlinearly through the shelf-life of stored RBCs and is linked to adverse transfusion outcomes as determined by chromium-51-labeled ($^{51}$Cr) posttransfusion recovery (PTR) studies in autologous donors (gold standard to determine RBC storage quality per European Council and US Food and Drug Administration guidelines).[11] Energy currencies such as ATP and NAD are also reduced during the storage period, the latter as a result of the activity of enzymes like ADP ribose hydrolases like CD38.[17] Accumulation of lipid peroxidation products and depletion of L-carnitine pools to repair such damage via the Lands cycle are also negative predictors of post-transfusion efficacy in mouse models and humans.[13,18]

Current blood bank quality control relies on hemolysis testing of only ~1% of monthly production via visual assessment and spectrophotometric methods.[19] Between assessment and transfusion, units may age in the refrigerator without individual evaluation, while storage lesions progress unpredictably due to donor-specific factor. This single-parameter approach inadequately correlates with posttransfusion recovery and effectiveness.[20,21] While the 51Cr-labeled PTR method is the gold standard for assessing stored blood quality and measuring transfusion outcomes, it remains prohibitively expensive due to radioactive handling requirements, specialized infrastructure, and costly disposal, limiting their clinical utility.[22] Emerging approaches include ultra-high throughput metabolomics, which has identified predictive biomarkers but has so far remained confined to research settings due to sophistication and cost. Functional assessment methods also show promise, including RBC morphology evaluation using scanning electron microscopy (SEM)[23] and differential interference contrast (DIC) microscopy,[24] and more recently, high-throughput imaging flow cytometry for single-cell classification.[2,25,26] Emerging photoacoustic microscopy[27] and Raman spectroscopy[28] techniques assess morphological changes of blood cells with potential for non-invasive clinical applications through blood storage bags or human skin. RBC deformability has also been recognized as a sensitive indicator of RBC functionality. Microsphiltration assays have recently been used to show that the sub-populations of RBCs that accumulate the bulk of the storage lesion are also the ones that are less deformable in vitro, and most likely to be sequestered in the spleen and erythrophagocytosed upon transfusion.[2,26] Deformability is commonly measured using micropore filtration,[29] micropipette aspiration,[30] and ektacytometry (e.g., LORRCA).[24,31] However, these approaches are inadequate for widespread clinical implementation due to high cost, specialized expertise requirements, and-in some cases-low throughput. Rapid, cost-effective, bedside technologies that correlate with metabolic profiles and predict transfusion efficacy are urgently needed to enable personalized transfusion medicine and improve clinical outcomes.[5]

[a]*Department of Mechanical Engineering, University of Colorado, Boulder, CO, 80309, USA. E-mail: Xiaoyun.Ding@colorado.edu*
[b]*Department of Biochemistry and Molecular Genetics, University of Colorado Anschutz Medical Campus, Aurora, CO, 80045, USA. E-mail: ANGELO.DALESSANDRO@CUANSCHUTZ.EDU*
[c]*Biomedical Engineering Program, University of Colorado, Boulder, CO, 80309, USA.*
[d]*Omix Technologies Inc, Aurora, CO, 80045, USA*
[e]*BioFrontiers Institute, University of Colorado, Boulder, CO 80309, USA.*
#These authors contributed equally.
†Electronic supplementary information (ESI) available. See DOI: 10.1039/x0xx00000x

To improve accessibility and throughput, lab-on-chip (LOC) technologies have emerged as promising tools for RBC quality assessment. These platforms offer portability, speed, and high-throughput capabilities. Many recent LOC techniques focus on miniaturizing traditional techniques, including imaging systems integrated with machine learning to profile the morphological heterogeneity of blood products with the goal to implement precision transfusion medicine practices.[32–35] Microfluidic adaptations of ektacytometry,[36] velocity-based deformability tracking,[37] ratchet-based cell sorting,[38] and capillary-mimicking deformability assays[39,40] have further enabled evaluations of RBC rigidity. However, these LOC platforms remain limited by sophisticated fabrication requirements and the need for skilled technical implementation.

Here, we present the Surface Acoustic Wave Hemolysis Assay (SAW-HA), an acoustic-integrated LOC platform that addresses these fabrication and operational challenges while introducing a new biophysical biomarker for assessing RBC storage quality. Compared to existing LOC methods, the SAW-HA offers several practical advantages: simple fabrication, minimal sample volume (<2 µL), rapid results (under two minutes), and no need for complex image analysis. SAW technologies are known for their precise fluid control and high biocompatibility.[41–45] Acoustic-induced heating has been used to precisely control on-chip temperatures,[46,47] with demonstrated applications in protein-ligand interaction screening and in distinguishing healthy from sickle cell disease samples.[48] Building on these capabilities, our assay leverages both acoustic forces and acoustic-induced heating to induce hemolysis, using the SAW Hemolysis Temperature (SAWHT) as a biomarker for storage-induced changes in RBC quality. Measuring SAWHT at weekly intervals during storage enables a quantitative assessment of RBC quality across a spectrum— from optimal to severely degraded units—supporting decisions on transfusion suitability. When integrated with metabolomics data, this platform provides new insights into donor-specific differences in stored RBC aging and introduces a biomarker to support the advancement of personalized transfusion medicine.

## Experimental methods

### Device fabrication

The SAW device was fabricated using a 500 µm thick, 76.2 mm diameter, 128° Y-cut X-propagating $LiNbO_3$ substrate. Two identical IDTs were patterned on either side of a PDMS microchannel. The IDTs were fabricated using standard photolithography, starting with spin-coating a positive photoresist (S1813, Dow, USA) on the wafer. After UV exposure and development with MF319 developer (Dow, USA), layers of chrome/gold (Cr/Au, 10/100 nm) were deposited using e-beam evaporation, and excess photoresist was removed via lift-off (Remover PG, Kayaku, Japan). Each IDT comprised 30 electrode pairs with 50 µm spacing and a 10 mm aperture, yielding a frequency of ≈20 MHz for the propagating SAW.

For the PDMS channel, a negative SU8 mold was prepared by spin-coating SU8 2025 photoresist (MicroChem, USA) on a silicon wafer and patterning it with optical lithography. PDMS (Sylgard 184, Dow Corning, USA) was poured onto the mold, cured at 65 °C for 35 minutes, and punched with 0.75 mm diameter inlet/outlet holes and a 0.35 mm diameter temperature measurement hole. The PDMS microchannel (80 µm height, 1 mm width, and 10 mm long) was bonded to the $LiNbO_3$ substrate using air plasma (PDC-001, Harrick Plasma, USA) and baked at 65 °C for 18 hours.

### Samples and materials

Leukodepleted pRBCs in ACD-A/AS-3 were collected via apheresis at Vitalant Blood Donation Center (Denver, CO, USA) from six healthy donors (3 males, 3 females, aged 35–40) with appropriate informed consent and institutional review board approval as specified in the Vitalant research materials agreement. RBC units were stored at 1–6 °C and aseptically sampled weekly from day 1 to day 42. Each week, 100 µL of RBC units were frozen for future metabolic and lipidomic analysis. For weekly experiments (SAW-HA, EHC, and temperature-matched SAW-HA), RBC units were diluted 1:10 in PBS (v/v).

### Device operation

For each set of experiments (n=3), the PDMS channel was filled with PBS, with additional PBS droplets placed at the inlet and outlet ports. The device was then placed in a mild vacuum chamber for 20 minutes under mild vacuum conditions for air bubble removal and temporary hydrophilicity treatment of the PDMS walls. Before each individual experiment, air bubbles were removed with ethanol, followed by manual injection of PBS into the inlet hole of the channel using a micropipette. Diluted RBC samples (~5 µL) were injected, ensuring the channel remained bubble-free. After each experiment, the channel was cleaned with bleach followed by water washing in preparation for the next experiment. The SAW device was mounted on a custom 3D-printed holder and imaged using an inverted microscope (Eclipse Ti2, Nikon, Japan), a CMOS camera (Orca-Flash 4.0, Hamamatsu, Japan), and HCImage Live software. Videos were recorded at 2.5 fps with a 10x objective in brightfield, capturing 2048x2048-pixel images at 50% brightness and minimum aperture. The same channel area was imaged for consistency. Temperature inside the PDMS channel was monitored using a digital thermocouple (5TC-TT-K-36-36, Newark, USA) connected to a data acquisition (DAQ) system (cDAQ-9171 and NI TB-9212, NI, USA) and controlled via LabVIEW (NI, USA), with measurements taken for each image frame. All thermocouples were calibrated using the National Instruments data acquisition system by measuring ice-bath temperature (0°C reference) to confirm measurement accuracy within ±0.01°C.

An RF signal generator (33500B, Keysight, USA) and two power amplifiers (403LA, E&I, USA) were used to apply signals to the IDTs via bayonet coupling adapter (BCA) cables to printed circuit board (PCB) connectors (Fig. S1A†). Each SAW device was individually characterized by measuring SAW resonant frequency using a network analyzer (E5061B, Keysight, USA). Resonant frequencies of the SAW devices were typically ranging from 19.56 to 19.58 MHz.

For SAW-HA, the SAW was operated at about 1.2 W for 2 minutes. For temperature-matched SAW-HA, the SAW power started at 0.8 W and manually increased by 0.2 W about every 10 s to align with the heating profile of EHC. For EHC, a transparent heating plate (BT-I55D, Cell MicroControls, Norfolk, VA, USA) with a microscope stage adapter (MSA-WELLP, Cell MicroControls, Norfolk, VA, USA) were used (Fig.S1B†). The temperature of the heating stage was controlled using a micro-temperature controller (mTC3-HT, Cell MicroControls, Norfolk, VA, USA) with the following parameters: band limit (Bnd) set to 100, final temperature set to 80°C, with all other parameters at factory defaults.

**Metabolomics/Lipidomics by UHPLC/MS**
To extract metabolites, either cold 5:3:2 MeOH:ACN:$H_2O$ (v/v/v) solution (metabolomics) or cold MeOH (lipidomics) was added in a 10:1 ratio to 5 µL of stored RBCs. Samples were vortexed vigorously for 30 minutes at 4°C, then centrifuged for 10 minutes at 18,213 rcf. Using 10 µL injection volumes, the supernatants were analyzed by ultra-high-pressure-liquid chromatography coupled to mass spectrometry (UHPLC-MS — Vanquish and Orbitrap Exploris 120, Thermo). Metabolites were resolved across a 1.7 µm, 2.1 x 150 mm Kinetex C18 column using a 5-minute gradient previously described.[49] Using 10 µL injection volumes, non-polar lipids were resolved using UHPLC coupled to $ddMS^2$ using a 5-minute gradient method as previously described.[50]

Following data acquisition, .raw files were converted to .mzXML using RawConverter. Metabolites were then annotated based on intact mass, $^{13}C$ natural isotope pattern and retention times in conjunction with the KEGG database and an in-house standard library. Peaks were integrated using El-Maven (Elucidata). Quality control was assessed as using technical replicates run at the beginning, end, and middle of each sequence as previously described. Lipidomics data were analyzed using LipidSearch 5.0 (Thermo Scientific), which provides lipid identification on the basis of accurate intact mass, isotopic pattern, and fragmentation pattern to determine lipid class and acyl chain composition.

Metabolomics and lipidomic data were analyzed using R Package (R Core Team) utilizing the following packages: shing, dplyr, plotly, ggplot2, circlize, and zip.

**Image capture and analysis**
All video and image processing were carried out using ImageJ[51]. For image analysis, a single rectangular region of interest (ROI) measuring 1400x2000 pixels was selected to cover the entire channel area. Cell lysis and protein denaturation curves were generated by utilizing the "Plot z-axis profile" function in ImageJ, which calculated the average grayscale intensity within the ROI for each frame throughout the image stack. The SAWHT of the RBCs was identified by aligning the rightmost peak of the curve with the corresponding temperature (refer to Fig. 1).

**Statistical analysis**
Statistical analyses were performed using Excel (Microsoft, USA) and OriginPro (OriginLab, USA) software. For comparisons of numerical means between datasets, a two-sided, unpaired student's t-test was used. Significance levels were set as follows: *$P < 0.05$, **$P < 0.01$, ***$P < 0.001$.

## Results and discussion
### Device design and working principle
The schematic and working principle of the SAW-HA are shown in Fig. 1. The system consists of a polydimethylsiloxane (PDMS) microfluidic channel bonded onto a piezoelectric lithium niobate ($LiNbO_3$) substrate. Two gold interdigital transducers (IDTs) are patterned symmetrically on either side of the 10 mm long, 80 µm high, and 1 mm wide channel. The device is placed in a custom-built holder fitted to a microscope stage and connected to BCA cables via a custom PCB connector (Fig. S1A†). Upon activation with radio frequency (RF) signal, the IDTs generate SAWs that propagate toward each other, forming a one-dimensional standing SAW field.[52] As the standing SAW interacts with the fluid inside the channel, acoustic energy radiates into the fluid due to the mismatch in sound velocity between the fluid and the substrate. This interaction induces pressure waves and acoustic heating effects caused by the viscous dissipation of acoustic energy into the fluid.[46] The pressure waves create alternating pressure nodes (regions of minimum pressure) and antinodes (regions of maximum pressure). Particles suspended in the fluid are driven toward these nodes or antinodes depending on acoustic contrast factors - their density and compressibility relative to the surrounding medium. The heating profile induced by the acoustic heating effect can be precisely tuned by altering parameters such as the standing SAW frequency, power input, duty cycle, and phase.

The standing SAW was applied to induce lysis of RBCs suspended in the microfluidic channel. The lysis process was tracked using high-resolution imaging and thermocouple-enabled temperature measurements, with grayscale intensity extracted from video data (Movie S1†). The intensity was plotted as a function of temperature to quantitatively represent the process (Fig. 1B). When the standing SAW was applied at 19.5 MHz and 1.2 W, RBCs aggregated at the pressure nodes due to acoustic radiation forces. This aggregation led to a noticeable increase in image grayscale intensity, which corresponded to the initial steep rise in the intensity-temperature graph. As the temperature increased from approximately 40 °C to 75 °C, the structural and mechanical properties of the RBC membranes began to change.[53] These changes included alterations in cell density, compressibility, and acoustic contrast factors. The intensity-temperature graph showed oscillations during this phase, which could potentially be attributed to morphological transitions in the cells. The peak in the graph at approximately 80 °C represents the point of maximum image intensity, corresponding to the rupture of RBC membranes. We define this temperature as the SAW Hemolysis Temperature (SAWHT), which serves as a biomarker for RBC storage quality assessment. Following membrane rupture, intracellular proteins (predominantly hemoglobin) undergo denaturation, aggregation, and precipitation, which is reflected in the declining portion of the curve.

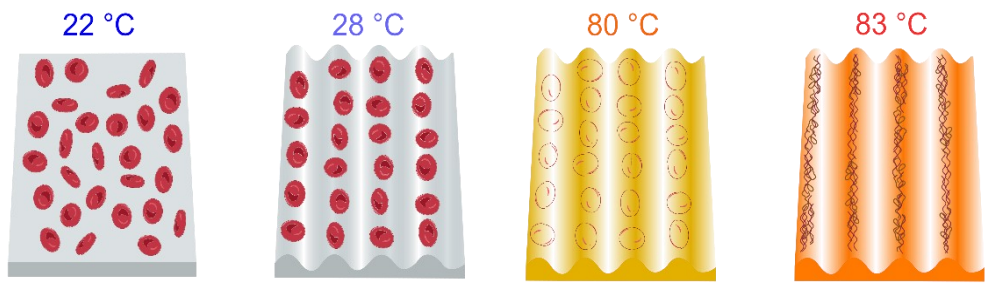

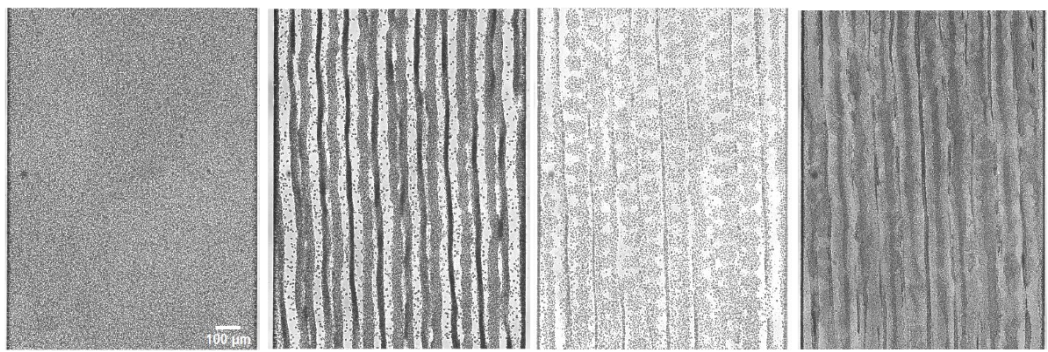

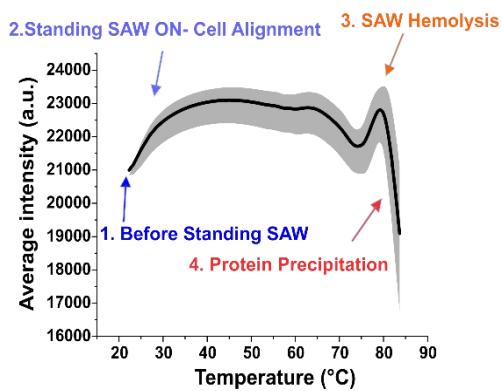

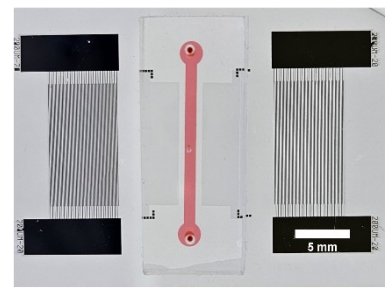

Fig. 1. Working mechanism of SAW Hemolysis Assay (SAW-HA). **(A)** (i) Schematic representation of the four sequential phases of RBC behavior in the microfluidic chamber under the standing SAW application: dispersed RBCs (22°C), cell alignment (28°C), RBC lysis (80°C), and protein aggregation (83°C). Cell and protein illustrations were created using BioRender. (ii) Representative high-contrast images correspond to the key phases outlined in (i). The scale bar represents 100 μm. Raw, unaltered images used for analysis are included in the supplementary materials (Fig. S2A). **(B)** Quantitative analysis of average grayscale intensity as a function of temperature from RBC samples of six donors using SAW-HA. The curve highlights distinct phases of cell behavior. Points corresponding to images in (A) are annotated. Note that the temperature at the third point corresponds to the biomarker SAW Hemolysis Temperature (SAWHT). **(C)** Photograph of the experimental microfluidic device used for the SAW-HA. The device features integrated SAW transducers and sample chamber with thermocouple probe hole in the middle, with a scale bar of 5 mm.

## SAWHT as a biomarker evaluating RBC storage quality change

We quantified the SAWHT of each donor's RBCs as a function of storage time (Fig. 2A). Two distinct trends are apparent. Four donors (Donors 1, 3, 4, and 6) maintained an unchanged acoustic signal over the 42-day storage period, with SAWHT showing minimal variation. In contrast, two donors (Donors 2 and 5) exhibited significant decreases in SAWHT. For the unchanged acoustic signal group (Donors 1, 3, 4, and 6), the average change in SAWHT over 42 days was -0.19 ± 0.61 °C. In contrast, the changed acoustic signal group (Donors 2 and 5) displayed significant declines in SAWHT starting at specific time points. Donor 2 showed a noticeable decrease from Day 28 onward, while Donor 5 experienced a similar reduction starting from Day 35. Both donors exhibited stabilization in SAWHT after the initial decrease, with no further significant changes observed through Day 42. The average decrease in SAWHT for this group was -2.12 ± 0.41 °C, significantly greater than that of the unchanged group (Fig. 2B).

It is well established that the melting temperatures of proteins are highly sensitive to heating rates, particularly in traditional thermal shift assays.[54,55] To ensure the reliability and comparability of results across storage periods and between donors, it is crucial to maintain a consistent heating rate during each trial. This approach minimizes variability arising from heating rate dependencies and allows for accurate assessment of stability changes in RBCs. Figure 2C illustrates the heating profiles of RBC samples obtained from six donors over a 42-day storage period. The profiles remained consistent across different weeks for individual donors, confirming that heating conditions were precisely controlled and uniform.

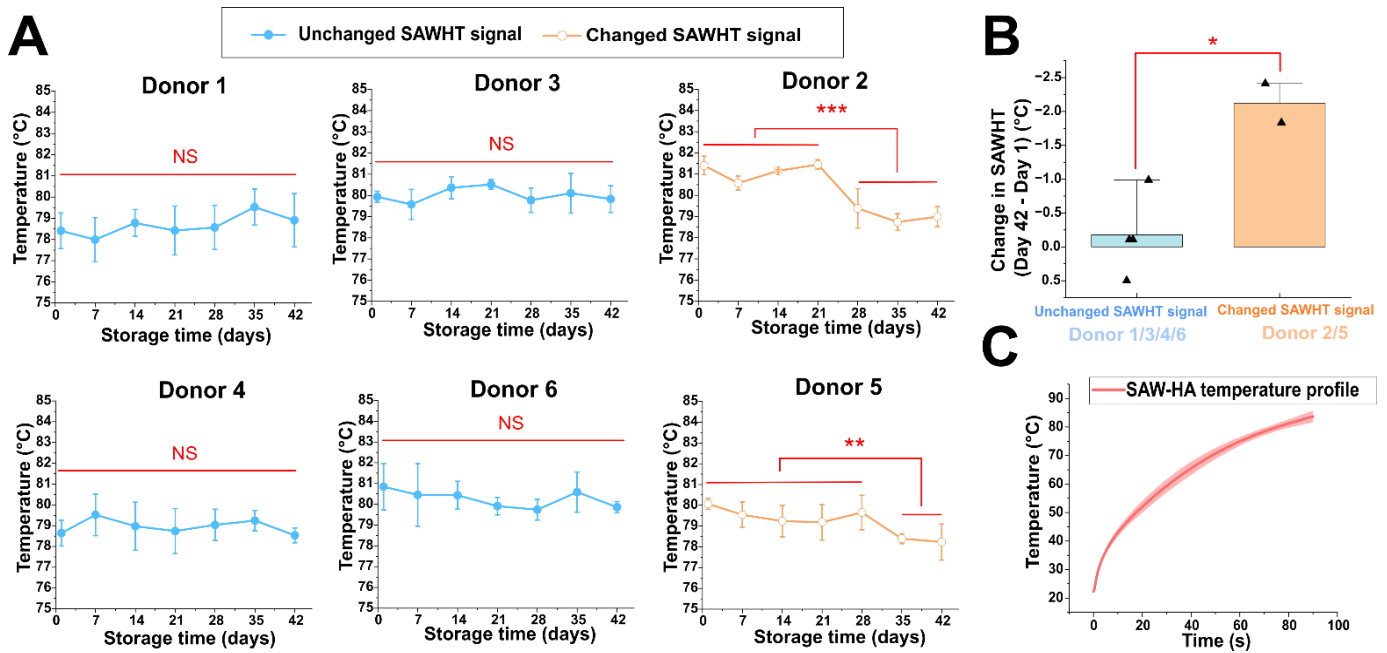

Fig. 2. SAW Hemolysis Temperature (SAWHT) serves as a biomarker for detecting donor-specific, storage-induced changes in RBC quality. **(A)** Weekly tracking of the SAWHT for each donor throughout 42 days of cold storage. Data points represent means ± s.d. from three independent trials per week for each donor. **(B)** Comparison of the change in SAWHT (Day 42 minus Day 1) for the two donor groups: unchanged acoustic signal (Donors 1, 3, 4, 6) and changed acoustic signal (Donors 2, 5). **(C)** The average temperature profile of the SAWHT across all six donors, demonstrating consistent temperature rise over time and its reproducibility. The shaded region represents the standard deviation. The student t-tests of independence were performed in the above figures. *$P < 0.05$, **$P < 0.01$, and ***$P < 0.001$; two-sided, unpaired t-test. NS, not significant.

## Acoustic effects play an important role in detecting RBC storage quality change

To isolate the role of acoustic waves beyond thermal contributions, we first conducted an electrical heating control (EHC) experiment. This control employed electrical heating under identical experimental conditions to the acoustic method, including channel dimensions, RBC concentrations, and video parameters (Fig. S1B†, Fig. S3†, & Movie S2†). The purpose of this experiment was to determine whether heating alone could account for the observed variations in SAWHT and to evaluate the effects of heating alone on detecting RBC storage quality change. The intensity versus temperature profiles obtained from the EHC (Fig. 3A) revealed distinct thermal transitions compared to the SAW-HA. An intensity decrease around 49 °C likely reflects morphological changes in RBCs due to denaturation of the cytoskeleton spectrin.[56] The rightmost peak in these profiles corresponds to the peak hemolysis point induced by electrical heating, which is followed by protein denaturation and precipitation. The average hemolysis temperature measured by the SAW-HA was significantly higher than that of the EHC (79.55 °C vs. 71.95 °C; Fig. 3D). Moreover, while donors 2 and 5 showed storage-dependent reductions in SAWHT using the SAW-HA, no such trend was observed in the EHC (Fig. 3E). This contrast highlights the importance of acoustic effects—beyond heating alone—in detecting storage-induced changes in RBC quality.

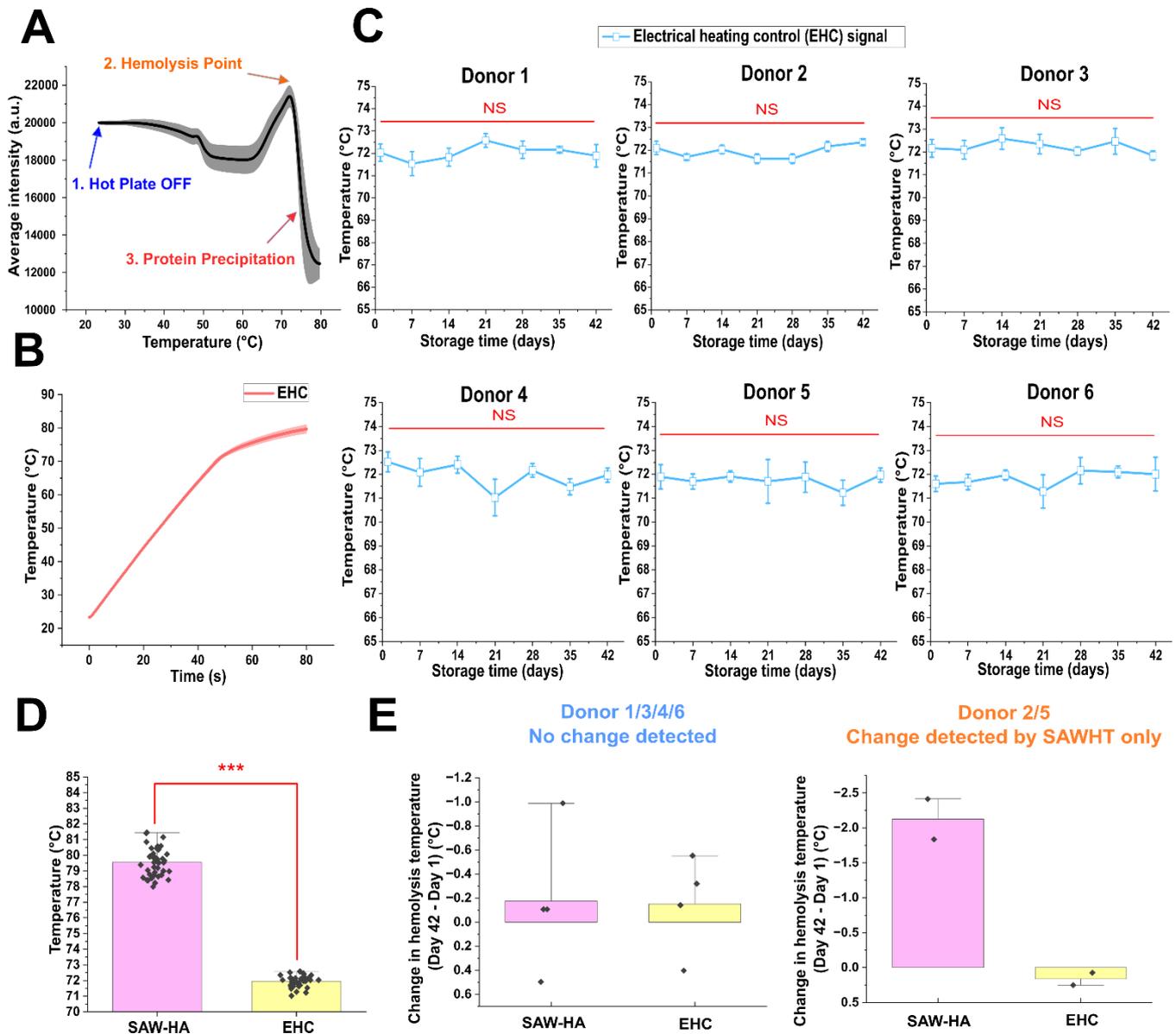

Fig. 3. Electrical heating control (EHC) shows that acoustic effects are critical for detecting changes of RBCs quality during cold storage. (A) Grayscale intensity versus temperature analysis of RBC samples from six donors under electrical heating reveals two key phases: (1) hemolysis point and (2) subsequent protein denaturation and precipitation. Representative images with enhanced contrast for these phases are included in Fig. S3. Raw, unaltered images used for analysis are included in Fig. S2B. (B) Average temperature profile of EHC from all six donors, demonstrating consistent and controlled temperature progression over time. The shaded region represents the standard deviation. (C) Weekly tracking of hemolysis temperatures across 42 days of cold storage for individual donors under EHC. Data points represent means ± s.d. from three independent trials per week for each donor. (D) Comparison of the hemolysis temperatures across all donors and weeks between SAW-HA and EHC. (E) Comparison of the change in hemolysis temperature (Day 42 minus Day 1) between SAW-HA and EHC for two donor groups: donors 1/3/4/6 (unchanged acoustic signal) and donors 2/5 (changed acoustic signal). The student t-test of independence was performed. ***P < 0.001; two-sided, unpaired t-test. NS, not significant.

To further confirm the role of acoustic waves and rule out potential confounding factors, we conducted an additional control experiment (Fig. S4†). In this experiment, the heating profile of the SAW-HA was adjusted to closely replicate that of the EHC (Fig. S4A†). By gradually increasing the acoustic power by 0.2 W every 10 seconds, we ensured that the temperature rise in this temperature-matched-to-EHC SAW-HA mirrored the slower heating profile observed in the EHC. The only key difference between this temperature-matched SAW-HA and the EHC is the presence of acoustic waves. The intensity versus temperature curves for the temperature-matched SAW-HA closely resembled those of the SAW-HA (Fig. S4C†). Both acoustic assays exhibited higher hemolysis temperatures (~80-82 °C) compared to the EHC (approximately 72 °C). Moreover, both assays detected storage-dependent shifts in hemolysis temperature for donors 2 and 5, a trend absent in the EHC. This finding demonstrates that the temperature difference in the initial of the heating profile or heat-shock effects (above 37 °C effect to cells) does not account for the observed donor-specific changes by the SAW-HA; rather, the presence of acoustic waves is essential for detecting these shifts.

**Metabolite correlations with SAWHT can differentiate storage duration and donor phenotypes in RBCs**

The relative abundance of metabolites central to energy and redox metabolism across all samples were correlated with storage duration (Fig. S5†). Similar to previous reports, we observed accumulations of RBC storage biomarkers hypoxanthine, 5-oxoproline, and lactate over the storage duration,[11,12] and glycolysis intermediates strongly negatively correlated with storage duration, in keeping with the body of literature on the metabolic changes of aging RBCs in vitro.[57] We first correlated the change in metabolite abundance with SAWHT across all storage days (Fig. 4A-B). The top positive correlates were several essential amino acids (leucine, glutamine) and energy metabolism substrates (AMP). Negative correlates include dicarboxylates (citrate, fumarate) and hydroxybutyrylcarnitine (acyl-C4:OH).

Next, we correlated the abundance of metabolites on day 1 with the SAWHT over storage duration to identify features that predict decreased lysis temperatures on day 42 (Fig. 4C-D). Here, positive correlates represent metabolites that stabilize SAWHT and include glycine, AMP, S-adenosylhomocysteine (SAH), and 8-methoxykynurenate. Negative correlates include ribose and biliverdin but are not as strongly correlated.

We then correlated the change in abundance of metabolites (Day 42 minus Day 1) with the SAWHT over storage duration (Fig. 4E-F). This analysis identified cysteine as the top negative correlate, with the two donor samples that showed decreased lysis temperature accumulating cysteine. Pyridoxal, an antioxidant and the active form of vitamin B6, is also strongly negatively correlated. Anthranilate, a tryptophan metabolite like kynurenine and marker of oxidative stress, was also negatively associated, as well as long-chain fatty acids (FAs). Allantoate, a purine metabolite, showed the strongest positive correlation. In this sense, positive correlation demonstrates metabolites that are decreased in donors with decreased lysis temperature. Citrate, a carboxylic acid that abounds in citrated-anticoagulants, was also identified positively correlated along with fumarate, a catabolite of purine salvage and malate/aspartate metabolism.[58,59]

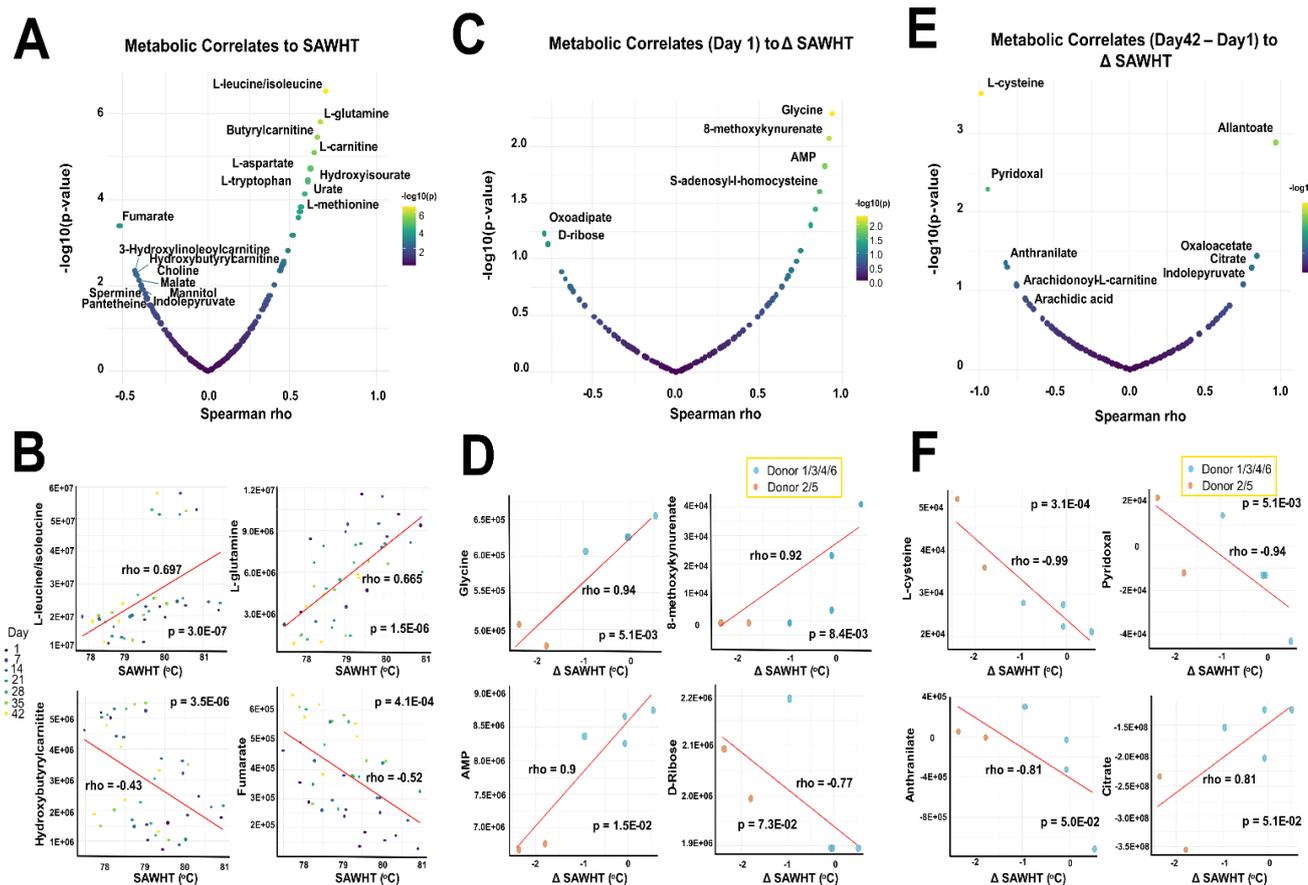

**Fig. 4.** A Spearman correlation analysis was performed between metabolite peak intensity (A.U.) and SAWHT. Metabolic correlates were graphed against p-value. Top correlates are labeled for **(A)** metabolite abundance versus SAWHT (N = 42), **(C)** Correlates of day 1 metabolite abundance to change in SAWHT over storage duration (Day 42 – Day 1, N = 6), and **(E)** Correlates of the change in metabolite abundance (Day 42 – Day 1) to changes in SAWHT over storage duration (Day 42 – Day 1, N = 6). **Selected metabolites were graphed (B)** for top correlates in (A), **(D)** for top correlates in (C), and **(F)** for top correlates in (E). For D and F, Donors 2 and 5 are marked in orange, with the remaining four donors in light blue.

To further interrogate the metabolic profiles during storage between the SAWHT-defined two groups of donors, we performed separate correlations of metabolite levels against change in SAWHT (Fig. 5A-B, Fig. S6†). These analyses revealed several key pathways that are selectively altered in donors with reduced SAWHT (Fig. 5C). Donors 2 and 5 showed loss of correlation in glutamine and its downstream metabolites such as GSH, while the other donors showed positive correlations and stabilization of SAWHT glutamate and GSH. Spermine, a critical polyamine that negatively correlates with SAWHT in other donors, shows a lack of correlation in donors 2 and 5. Of major importance, tryptophan metabolism is drastically altered across multiple nodes in donors 2 and 5, with kynurenine, anthranilate, and serotonin showing reduced abundance and association with SAWHT. Furthermore, these donors show elevated TG concentration across all temperatures. Together, these results suggest that BMI may be tightly correlated with the SAWHT depression over time, as elevated TGs and disrupted tryptophan metabolism are associated with high BMI.[60–62]

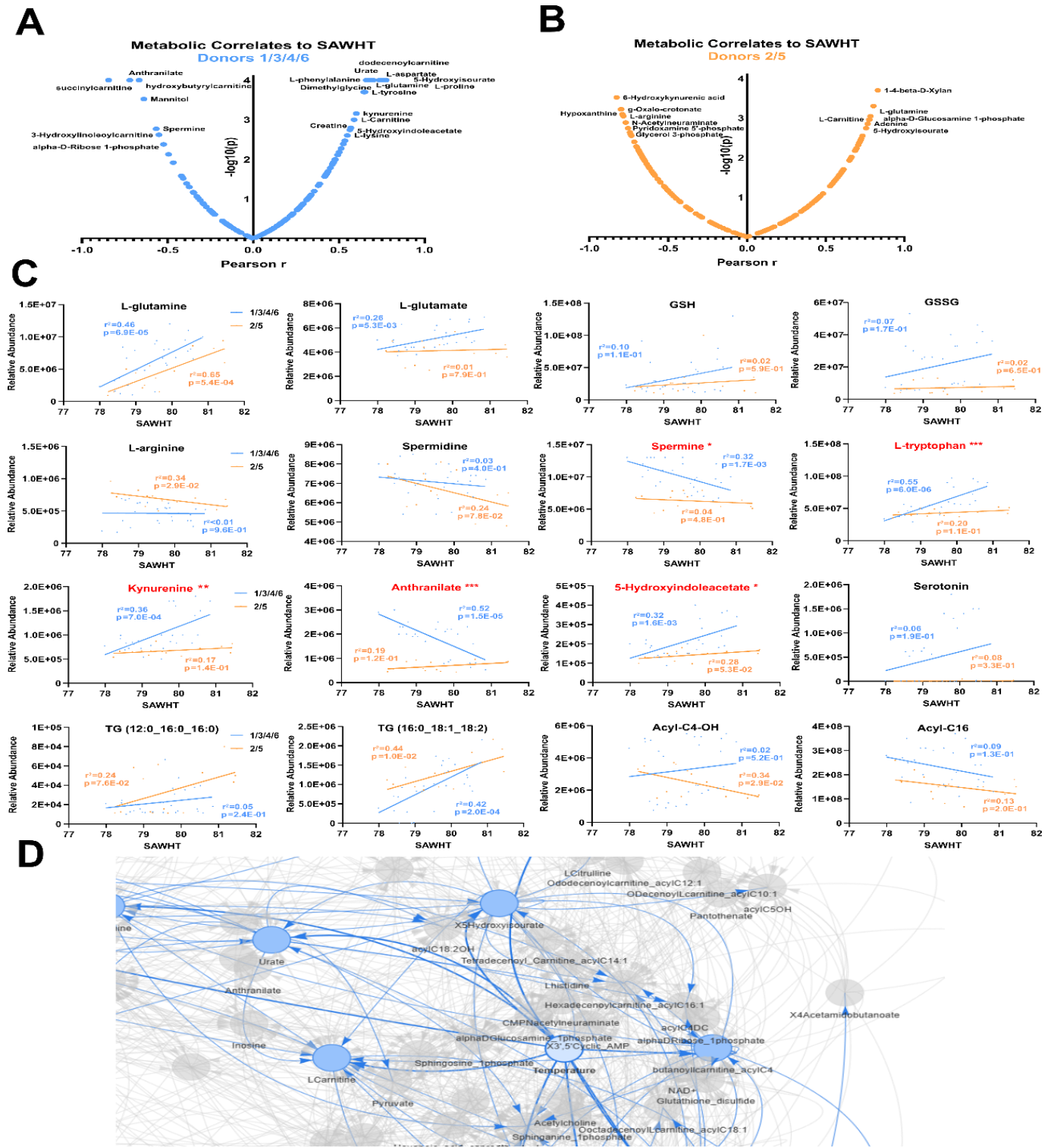

**Fig. 5.** A Pearson correlation analysis was independently performed between metabolite/lipid peak intensity (A.U.) and SAWHT within the two donor groups. Top correlates are labeled for **(A)** Donors 1,3,4, and 6 (N = 28) and **(B)** Donors 2 and 5 (N = 14). **(C)** Select features were graphed with significance marked for features with significantly different slopes between two donor groups by simple linear regression ($p < 0.05$ *, $p < 0.01$ **, $p < 0.001$ ***). **(D)** A neural network analysis centered around metabolites that correlate to SAWHT (N = 42).

Current FDA guidelines permit refrigerated RBC storage up to 42 days, requiring a mean 24-hour autologous transfusion recovery rate of at least 75% across tested samples.[22] However, statistical modeling of existing recovery data suggests only 67.3% of blood unit samples would meet this standard at 42 days, revealing significant variability in stored RBC quality that remains undetected in clinical practice due to the high cost and specialized expertise required for PTR testing.[63] A major contributor is donor-to-donor differences, with some individuals consistently producing better-storing RBCs than others.[64] Factors impacting the heterogeneity in the quality and post-transfusion efficacy of stored blood products include genetics,[8,65] biology (sex, age, ethnicity),[6] BMI,[7] and recreational, professional, environmental, or other exposures.[66] Such variability poses significant risks for transfusion-dependent patients—such as those with thalassemia, sickle cell anemia, or trauma—who require frequent or massive transfusions. RBC units from poor storers can undergo rapid clearance, leading to iron overload,[4] which in turn promotes inflammatory complications such as acute respiratory distress syndrome and cardiorenal dysfunction[67,68] and increases susceptibility to infections by siderophilic bacteria.[69] Identifying high-quality RBC units based on donor characteristics may improve transfusion strategies, patient outcomes, and reduce healthcare costs.

Several technologies have been developed to assess RBC quality and reveal donor-donor differences, including metabolic biomarkers,[7,13,64,70] morphology analysis,[25,26,32,33] and deformability-based microfluidic tests.[37,38] However, these methods require large and sophisticated equipment, complex operation, and in some cases, lengthy processing times, limiting their feasibility for large-scale blood quality assessment. Our acoustic-integrated microfluidic device, the SAW Hemolysis Assay, offers a simple, rapid, and cost-effective solution. It directly analyzes unpurified pRBCs without further sample processing, requires only 2 μL of blood, and provides results within two minutes.

One of the key innovations of our method is the introduction of the SAW Hemolysis Temperature as a biomarker for RBC storage quality. Unlike traditional assays where lysis is a preprocessing step for intracellular content analysis,[71] our results demonstrate that the temperature at which RBCs lyse by acoustics provides valuable insights into membrane integrity and biomechanical resilience. To enable widespread clinical adoption, with technological advancement, there is significant potential for SAW-HA integration into a miniaturized PCB-based device supporting simultaneous operation of multiple SAW-HA units with on-board power management for portable deployment. We envision this multi-device platform addressing throughput requirements for routine blood quality monitoring, while sterile sampling innovations using interconnected primary and aliquot bag systems could enable representative testing without compromising transfusable unit integrity. Combined with standard photolithography fabrication for scalable manufacturing, this integrated system could transform SAW-HA into a clinically viable platform for high-throughput blood product quality assessment in blood banking applications.

After confirming here that donor RBCs stored in a manner consistent with the literature (Fig. S5†), our correlation analyses identified several SAW-specific metabolites associated with major changes in SAWHT (Fig. 4A-B). Metabolites strongly positively correlated to SAWHT represent potential biomarkers for healthy membrane stability. This list includes many amino acids, each contributing to a multitude of vital cellular functions. For instance, glutamine is a conditionally essential amino acid and player in erythropoiesis, as well as the precursor to the vital endogenous antioxidant glutathione.[72,73] Tryptophan, the amino acid precursor to kynurenine and serotonin, was also observed to maintain healthy membrane stability.[74] Coincidentally, tryptophan metabolism has been shown to be dysregulated in obesity and metabolic dysfunction across multiple tissues.[75–77] Additionally, the tryptophan metabolite kynurenine is a top metabolic correlate to osmotic stress and hemolysis despite a lack of storage-mediated changes.[12] In the realm of fatty acid metabolism, membrane stabilizer carnitine and hydroxylated acyl-carnitines were strong negative correlates, a potential double-hit to FA equilibrium.[78]

We identified several metabolites as potential predictors of decreased lysis temperature over the storage duration (Fig. 4C-D) including ribose and oxoadipate. While oxoadipate is an intermediate of tryptophan metabolism, D-ribose is a protein-glycation substrate, another critical process for membrane proteins as well as a player in oxidative stress and energy metabolism.[79] Inversely, we identified a few potential predictors of stable lysis temperature over the storage duration. AMP was decreased in the donor samples with reduced SAWHT. ATP is associated with erythrocyte membrane response to high pressure through alterations in phosphorylation and dephosphorylation of membrane proteins.[80] Glycine was the top correlated metabolite, most likely due to its role in glutathione synthesis and regulation of oxidative signalling and damage.[81]

Finally, we highlight metabolites that correlate with changes in SAWHT (Fig. 4E-F). Cysteine shows the strongest negative correlation, with the two donors with decreased SAWHT accumulating cysteine over the storage duration. While cysteine has been implicated in blood fluidity in vitro, it is most likely a marker of impaired glutathione synthesis.[82,83] Thus, it is unsurprising that features involved in redox balance are also implicated, such as pyridoxal, anthranilate, and allantoate. Pyridoxal, an aldehyde of vitamin B6, is implicated in FA synthesis and acts as an ROS scavenger.[84] Anthranilate is another tryptophan metabolite and observed marker of donor age in stored RBCs.[85] Purine metabolite allantoate is also a biomarker of donor age and an established marker of oxidative stress.[86]

Separation of the donor groups based on SAWHT over storage duration reveals that tryptophan and arginine metabolism over time are significantly altered over several nodes (Fig. 5A-C). Additionally, glutamate, glutathione, and glutathione disulfide show downward trends. These pathways appear to be the most influential when predicting SAW hemolysis performance. A neural network analysis further identified L-carnitine, pyruvate and hydroxylated acyl-carnitines as associated players (Fig. 5D). Urate metabolism was identified as a nearby node to SAWHT, which has been shown to be associated with high BMI and increased hemolysis.[87]

Although these pathways are important in blood aging, donor statistics also play an important role in the stability of the blood sample. Donors 2 and 5 have the highest BMI among the group (Table 1). There is a significant overlap between the biomarkers associated with hemolysis and those associated with high BMI. For instance, the high correlation of SAWHT with abundance of leucine and isoleucine could indicate insulin resistance associated with high BMI.[88] Additionally, triglycerides (TGs) are among the lipid correlates to SAWHT (Fig. S6B† and Fig. 5C). Elevated TGs are not only observed in obesity and associated with high BMI but are also a critical component of the RBC membrane.[60,61] Upon further review of donor statistics, it was discovered that donors 2 and 5 were of Hispanic heritage (Table 1). Ethnicity has been shown to be a significant factor in RBC membrane-lipid profile, median BMI, and hemolysis in larger population studies. One such study in over 16,000 US Hispanic/Latinos found an overall prevalence of dyslipidemia in Hispanics at 65% and noted high incidences of elevated low-density lipoprotein cholesterol (LDL-C) and triglycerides[89]. When evaluating ethnicity as a variable in the determination of insulin resistance via triglyceride levels, Mexican Americans had higher prevalence of syndromes associated with insulin resistance than non-Hispanic blacks despite similar occurrence of obesity, hypertension, and diabetes between the two groups [90]. Of interest to our SAW-HA, Hispanic blood donors exhibited higher markers of oxidative hemolysis than other donor populations [91]. However, we report an altered metabolic profile for this SAW hemolysis when compared to either oxidative or osmotic hemolysis markers previously observed [12]. Thus, with the interplay of ethnicity, BMI, lipid profile, and hemolysis; we cannot single out one factor as the main contributor to our observed decrease in SAW-hemolysis temperature. Rather, we can conclude that these factors combine to alter membrane stability during storage, and our method can identify donor samples that may have decreased efficacy as transfusion medicine. Finally, deficient tryptophan metabolism and reduced serotonin are also associated with high BMI, potentially explaining the lack of correlation in our donor set.[62,75]

Table 1. Donor demographic information. Donors are arranged in order of increasing BMI. Note that donors 2 and 5 exhibit the highest BMI values (36.9 and 33.0 lbs/in² respectively) color, classified as Class II and Class I obesity according to CDC guidelines [60].

| Donor | Sex | Race/ethnicity | Hemoglobin concentration (g/dL) | Cholesterol (mg/dL) | Height | Weight (lbs) | BMI (lbs/in²) |
|---|---|---|---|---|---|---|---|
| 3 | M | White | 18.2 | 155 | 5'10'' | 194 | 27.8 |
| 1 | M | White | 14.8 | 192 | 6'1'' | 214 | 28.2 |
| 4 | F | White | 14.7 | 145 | 5'7'' | 182 | 28.5 |
| 6 | F | White | 13.5 | 134 | 5'7'' | 197 | 30.9 |
| 5 | F | Other Hispanic | 14.7 | 179 | 5'7'' | 211 | 33 |
| 2 | M | Puerto Rican | 17.3 | 143 | 5'9'' | 250 | 36.9 |

The SAW-HA provides a uniquely sensitive method for evaluating subtle changes in RBC membrane stability during cold storage. The mechanisms underlying this capability can be explained by the complex interactions between acoustic forces and cellular structures. Standing SAW imposes cyclic mechanical stresses (shear stresses, coupling with rapidly changing acoustic pressure fields and radiation pressures[47]) that actively probe membrane fragility, revealing donor-specific mechanical vulnerabilities masked under thermal conditions alone. Our data demonstrates that acoustic methods produce significantly higher hemolysis temperatures than heat-only controls (79.55 °C vs. 71.95 °C), suggesting that acoustic forces fundamentally alter the cells' response to thermal stress. This effect likely occurs through multiple pathways: 1) acoustic forces may transiently reorganize lipid packing, as molecular dynamics simulations show that ultrasound can induce oscillatory pressure changes that alter lipid tail ordering;[92,93] 2) mechanical stimulation may modify membrane protein conformations through hydrophobic mismatch mechanisms, where small changes in bilayer thickness alters proteins toward new functional states;[94] and 3) acoustic perturbations likely engage the spectrin-actin cytoskeletal network, potentially strengthening membrane-skeleton coupling through mechanosensitive channel activation (e.g., Piezo1) and calcium-dependent pathways.[95]

Our data highlights several metabolic correlates with the new SAWHT metric described. However, high BMI and the associated metabolic phenotype seem to be an influential factor in our limited donor group. Still, the LOC protocol identified donor samples with decreased membrane stability, thus predicted to have decreased transfusion efficacy. As the transfusion field moves toward personalized medicine, assessing donor-specific factors such as BMI may be critical for optimizing blood storage and recipient outcomes. Our SAW-integrated device provides a scalable, rapid, and accessible approach to RBC quality assessment, offering a promising on-site tool for improving transfusion practices while addressing critical concerns about donor variability. Future objectives include larger scale, follow-up studies in combination with current non-bedside assays of transfusion efficacy.

## Conclusions

In summary, we have developed the Surface Acoustic Wave Hemolysis Assay (SAW-HA), a microfluidic platform that introduces SAW Hemolysis Temperature (SAWHT) as a quantitative biomarker for evaluating stored red blood cell quality. The SAW-HA differentiated donor-specific storage behaviours across six donors over 42 days, identifying two distinct phenotypes: donors with stable SAWHT (-0.19 ± 0.61 °C) and those with declining SAWHT (-2.12 ± 0.41 °C). Integration with metabolomics revealed that donors with decreased SAWHT were characterized by disrupted redox control, deficient tryptophan metabolism, and elevated triglycerides—metabolic biomarkers indicative of decreased membrane stability—with high BMI emerging as a potential predictor of poor storage quality. Even in such a small sample group, SAWHT measurement was able to separate donor samples displaying markers of decreased membrane stability independently of donor ethnicity or metabolomics profile. This proof-of-concept platform addresses critical limitations in transfusion medicine by providing rapid (<2 minutes), sample-efficient (<2 µL), and cost-effective point-of-care analysis. It should, however, be noted that clinical translation requires validation through larger studies correlating SAWHT with gold-standard post-transfusion recovery methods, and metabolic correlations need confirmation across diverse populations to establish broader applicability. We believe the SAW-HA platform holds significant promise for advancing personalized transfusion medicine through rapid bedside quality assessment. Future applications may extend to diagnosing blood diseases, detecting pathological conditions that alter cellular membranes, and conducting fundamental membrane biophysics studies, potentially revolutionizing blood banking practices and improving patient outcomes.

## Author contributions

M.S., C.C.A., A.D., and X.D., designed the study. C.C.A., J.A.R., A.D. performed omics analyses. X.D. contributed to the conceptualization. M.S., N.S., performed acoustic shift assays. M.S., C.C.A. wrote the first draft of the manuscript, which was reviewed and finalized by all co-authors.

#These authors contributed equally.
* Corresponding authors:
Xiaoyun.Ding@colorado.edu,
Angelo.Dalessandro@cuanschutz.edu

## Conflicts of interest
A patent application based on this work is filed.

## Data availability

All data supporting the findings of this study are included in the main text and the ESI. † The omics raw data is available at the NIH Common Fund's National Metabolomics Data Repository (NMDR) website, the Metabolomics Workbench, https://www.metabolomicsworkbench.org where it has been assigned Project ID ST004148. The data can be accessed directly via it's Project DOI: 10.21228/M88R80 . This work is supported by Metabolomics Workbench/National Metabolomics Data Repository (NMDR) (grant# U2C-DK119886), Common Fund Data Ecosystem (CFDE) (grant# 3OT2OD030544) and Metabolomics Consortium Coordinating Center (M3C) (grant# 1U2C-DK119889).

## Acknowledgements

The microfluidic devices were fabricated in JILA clean room at University of Colorado Boulder. We acknowledge funding support from NIH MIRA Award 1R35GM142817 and W. M. Keck Foundation grant to X.D..

## Notes and references

1. Zimring JC. Established and theoretical factors to consider in assessing the red cell storage lesion. Blood. 2015;125:2185–90.
2. Peltier S, Marin M, Dzieciatkowska M, Dussiot M, Roy MK, Bruce J, et al. Proteostasis and metabolic dysfunction characterize a subset of storage-induced senescent erythrocytes targeted for post-transfusion clearance. J Clin Invest [Internet]. 2025 [cited 2025 Apr 29]; Available from: https://www.jci.org/articles/view/183099
3. Rabcuka J, Blonski S, Meli A, Sowemimo-Coker S, Zaremba D, Stephenson D, et al. Metabolic reprogramming under hypoxic storage preserves faster oxygen unloading from stored red blood cells. Blood Adv. 2022;6:5415–28.
4. Hod EA, Zhang N, Sokol SA, Wojczyk BS, Francis RO, Ansaldi D, et al. Transfusion of red blood cells after prolonged storage produces harmful effects that are mediated by iron and inflammation. Blood. 2010;115:4284–92.
5. Isiksacan Z, D'Alessandro A, Wolf SM, McKenna DH, Tessier SN, Kucukal E, et al. Assessment of stored red blood cells through lab-on-a-chip technologies for precision transfusion medicine. Proc Natl Acad Sci. 2023;120:e2115616120.
6. Kanias T, Lanteri MC, Page GP, Guo Y, Endres SM, Stone M, et al. Ethnicity, sex, and age are determinants of red blood cell storage and stress hemolysis: results of the REDS-III RBC-Omics study. Blood Adv. 2017;1:1132–41.
7. Hazegh K, Fang F, Bravo MD, Tran JQ, Muench MO, Jackman RP, et al. Blood donor obesity is associated with changes in red blood cell metabolism and susceptibility to hemolysis in cold storage and in response to osmotic and oxidative stress. Transfusion (Paris). 2021;61:435–48.
8. Page GP, Kanias T, Guo YJ, Lanteri MC, Zhang X, Mast AE, et al. Multiple-ancestry genome-wide association study identifies 27 loci associated with measures of hemolysis following blood storage. J Clin Invest [Internet]. 2021 [cited 2024 Dec 31];131. Available from: https://www.jci.org/articles/view/146077
9. D'Alessandro A, Hod EA. Red Blood Cell Storage: From Genome to Exposome towards Personalized Transfusion Medicine. Transfus Med Rev. 2023;37:150750.
10. Koch CG, Duncan AI, Figueroa P, Dai L, Sessler DI, Frank SM, et al. Real Age: Red Blood Cell Aging During Storage. Ann Thorac Surg. 2019;107:973–80.
11. Paglia G, D'Alessandro A, Rolfsson Ó, Sigurjónsson ÓE, Bordbar A, Palsson S, et al. Biomarkers defining the metabolic age of red blood cells during cold storage. Blood. 2016;128:e43-50.
12. Nemkov T, Stephenson D, Erickson C, Dzieciatkowska M, Key A, Moore A, et al. Regulation of kynurenine metabolism by blood donor genetics and biology impacts red cell hemolysis in vitro and in vivo. Blood. 2024;143:456–72.
13. Nemkov T, Key A, Stephenson D, Earley EJ, Keele GR, Hay A, et al. Genetic regulation of carnitine metabolism controls lipid damage repair and aging RBC hemolysis in vivo and in vitro. Blood. 2024;143:2517–33.
14. Fenk S, Melnikova EV, Anashkina AA, Poluektov YM, Zaripov PI, Mitkevich VA, et al. Hemoglobin is an oxygen-dependent glutathione buffer adapting the intracellular reduced glutathione levels to oxygen availability. Redox Biol. 2022;58:102535.
15. Keele GR, Dzieciatkowska M, Hay AM, Vincent M, O'Connor C, Stephenson D, et al. Genetic architecture of the red blood

cell proteome in genetically diverse mice reveals central role of hemoglobin beta cysteine redox status in maintaining circulating glutathione pools [Internet]. bioRxiv; 2025 [cited 2025 Apr 29]. p. 2025.02.27.640676. Available from: https://www.biorxiv.org/content/10.1101/2025.02.27.640676v1
16. Whillier S, Raftos JE, Sparrow RL, Kuchel PW. The effects of long-term storage of human red blood cells on the glutathione synthesis rate and steady-state concentration. Transfusion (Paris). 2011;51:1450–9.
17. Nemkov T, Stephenson D, Earley EJ, Keele GR, Hay A, Key A, et al. Biological and genetic determinants of glycolysis: Phosphofructokinase isoforms boost energy status of stored red blood cells and transfusion outcomes. Cell Metab. 2024;36:1979-1997.e13.
18. D'Alessandro A, Keele GR, Hay A, Nemkov T, Earley EJ, Stephenson D, et al. Ferroptosis regulates hemolysis in stored murine and human red blood cells. Blood. 2025;145:765–83.
19. Hess JR, Sparrow RL, Van Der Meer PF, Acker JP, Cardigan RA, Devine DV. BLOOD COMPONENTS: Red blood cell hemolysis during blood bank storage: using national quality management data to answer basic scientific questions: RBC HEMOLYSIS DURING STORAGE. Transfusion (Paris). 2009;49:2599–603.
20. L'Acqua C, Bandyopadhyay S, Francis RO, McMahon DJ, Nellis M, Sheth S, et al. Red blood cell transfusion is associated with increased hemolysis and an acute phase response in a subset of critically ill children. Am J Hematol. 2015;90:915–20.
21. Rapido F, Brittenham GM, Bandyopadhyay S, La Carpia F, L'Acqua C, McMahon DJ, et al. Prolonged red cell storage before transfusion increases extravascular hemolysis. J Clin Invest. 127:375–82.
22. Roussel C, Buffet PA, Amireault P. Measuring Post-transfusion Recovery and Survival of Red Blood Cells: Strengths and Weaknesses of Chromium-51 Labeling and Alternative Methods. Front Med. 2018;5:130.
23. Mustafa I, Al Marwani A, Mamdouh Nasr K, Abdulla Kano N, Hadwan T. Time Dependent Assessment of Morphological Changes: Leukodepleted Packed Red Blood Cells Stored in SAGM. BioMed Res Int. 2016;2016:4529434.
24. Marin M, Roussel C, Dussiot M, Ndour PA, Hermine O, Colin Y, et al. Metabolic rejuvenation upgrades circulatory functions of red blood cells stored under blood bank conditions. Transfusion (Paris). 2021;61:903–18.
25. Roussel C, Dussiot M, Marin M, Morel A, Ndour PA, Duez J, et al. Spherocytic shift of red blood cells during storage provides a quantitative whole cell–based marker of the storage lesion. Transfusion (Paris). 2017;57:1007–18.
26. Roussel C, Morel A, Dussiot M, Marin M, Colard M, Fricot-Monsinjon A, et al. Rapid clearance of storage-induced microerythrocytes alters transfusion recovery. Blood. 2021;137:2285–98.
27. Strohm EM, Berndl ESL, Kolios MC. Probing Red Blood Cell Morphology Using High-Frequency Photoacoustics. Biophys J. 2013;105:59–67.
28. Atkins CG, Schulze HG, Chen D, Devine DV, Blades MW, Turner RFB. Using Raman spectroscopy to assess hemoglobin oxygenation in red blood cell concentrate: an objective proxy for morphological index to gauge the quality of stored blood? Analyst. 2017;142:2199–210.
29. Berezina TL, Zaets SB, Morgan C, Spillert CR, Kamiyama M, Spolarics Z, et al. Influence of Storage on Red Blood Cell Rheological Properties. J Surg Res. 2002;102:6–12.
30. Paulitschke M, Nash GB. Micropipette methods for analysing blood cell rheology and their application to clinical research. Clin Hemorheol Microcirc. 1993;13:407–34.
31. van Cromvoirt AM, Fenk S, Sadafi A, Melnikova EV, Lagutkin DA, Dey K, et al. Donor Age and Red Cell Age Contribute to the Variance in Lorrca Indices in Healthy Donors for Next Generation Ektacytometry: A Pilot Study. Front Physiol [Internet]. 2021 [cited 2025 Mar 20];12. Available from: https://www.frontiersin.org/journals/physiology/articles/10.3389/fphys.2021.639722/full
32. Sierra F. DA, Melzak KA, Janetzko K, Klüter H, Suhr H, Bieback K, et al. Flow morphometry to assess the red blood cell storage lesion. Cytometry A. 2017;91:874–82.
33. Recktenwald SM, Lopes MGM, Peter S, Hof S, Simionato G, Peikert K, et al. Erysense, a Lab-on-a-Chip-Based Point-of-Care Device to Evaluate Red Blood Cell Flow Properties With Multiple Clinical Applications. Front Physiol [Internet]. 2022 [cited 2025 Apr 10];13. Available from: https://www.frontiersin.org/journals/physiology/articles/10.3389/fphys.2022.884690/full
34. Yao Y, He L, Mei L, Weng Y, Huang J, Wei S, et al. Cell damage evaluation by intelligent imaging flow cytometry. Cytometry A. 2023;103:646–54.
35. Kaestner L, Schlenke P, von Lindern M, El Nemer W. Translatable tool to quantitatively assess the quality of red blood cell units and tailored cultured red blood cells for transfusion. Proc Natl Acad Sci. 2024;121:e2318762121.
36. Xu Z, Zheng Y, Wang X, Shehata N, Wang C, Sun Y. Stiffness increase of red blood cells during storage. Microsyst Nanoeng. 2018;4:17103.
37. Guruprasad P, Mannino RG, Caruso C, Zhang H, Josephson CD, Roback JD, et al. Integrated automated particle tracking microfluidic enables high-throughput cell deformability cytometry for red cell disorders. Am J Hematol. 2019;94:189–99.
38. Islamzada E, Matthews K, Guo Q, Santoso AT, Duffy SP, Scott MD, et al. Deformability based sorting of stored red blood cells reveals donor-dependent aging curves. Lab Chip. 2020;20:226–35.
39. Zheng Y, Chen J, Cui T, Shehata N, Wang C, Sun Y. Characterization of red blood cell deformability change during blood storage. Lab Chip. 2013;14:577–83.
40. Man Y, Kucukal E, An R, Watson QD, Bosch J, Zimmerman PA, et al. Microfluidic assessment of red blood cell mediated microvascular occlusion. Lab Chip. 2020;20:2086–99.
41. Bhadra J, Sridhar N, Fajrial AK, Hammond N, Xue D, Ding X. Acoustic streaming enabled moderate swimming exercise reduces neurodegeneration in *C. elegans*. Sci Adv. 2023;9:eadf5056.
42. Gao Y, Voglhuber-Brunnmaier T, Li Y, Akh L, Patino NH, Fajrial AK, et al. Reconfiguring Surface Acoustic Wave Microfluidics via In Situ Control of Elastic Wave Polarization. Phys Rev Lett. 2025;134:037002.
43. Shakya G, Yang T, Gao Y, Fajrial AK, Li B, Ruzzene M, et al. Acoustically manipulating internal structure of disk-in-sphere endoskeletal droplets. Nat Commun. 2022;13:987.
44. Gao Y, Liu K, Lakerveld R, Ding X. Staged Assembly of Colloids Using DNA and Acoustofluidics. Nano Lett. 2022;22:6907–15.

45. Sridhar N, Fajrial AK, Doser R, Hoerndli F, Ding X. Surface Acoustic Wave Microfluidics for Repetitive and Reversible Temporary Immobilization of C. elegans. Lab Chip. 2022;10.1039.D2LC00737A.
46. Shilton RJ, Mattoli V, Travagliati M, Agostini M, Desii A, Beltram F, et al. Rapid and Controllable Digital Microfluidic Heating by Surface Acoustic Waves. Adv Funct Mater. 2015;25:5895–901.
47. Reboud J, Bourquin Y, Wilson R, Pall GS, Jiwaji M, Pitt AR, et al. Shaping acoustic fields as a toolset for microfluidic manipulations in diagnostic technologies. Proc Natl Acad Sci. 2012;109:15162–7.
48. Ding Y, Ball KA, Webb KJ, Gao Y, D'Alessandro A, Old WM, et al. On-Chip Acousto Thermal Shift Assay for Rapid and Sensitive Assessment of Protein Thermodynamic Stability. Small. 2020;16:2003506.
49. Nemkov T, Reisz JA, Gehrke S, Hansen KC, D'Alessandro A. High-Throughput Metabolomics: Isocratic and Gradient Mass Spectrometry-Based Methods. In: D'Alessandro A, editor. High-Throughput Metabolomics: Methods and Protocols [Internet]. New York, NY: Springer; 2019 [cited 2025 Apr 29]. p. 13–26. Available from: https://doi.org/10.1007/978-1-4939-9236-2_2
50. Reisz JA, Dzieciatkowska M, Stephenson D, Gamboni F, Morton DH, D'Alessandro A. Red Blood Cells from Individuals with Lesch–Nyhan Syndrome: Multi-Omics Insights into a Novel S162N Mutation Causing Hypoxanthine-Guanine Phosphoribosyltransferase Deficiency. Antioxidants. 2023;12:1699.
51. Schneider CA, Rasband WS, Eliceiri KW. NIH Image to ImageJ: 25 years of image analysis. Nat Methods. 2012;9:671–5.
52. Gao Y, Fajrial AK, Yang T, Ding X. Emerging on-chip surface acoustic wave technology for small biomaterials manipulation and characterization. Biomater Sci. 2021;9:1574–82.
53. Christel S, Little C. Morphological changes during heating of erythrocytes from stored human blood. J Therm Biol. 1984;9:221–8.
54. Toda A. Heating rate dependence of melting peak temperature examined by DSC of heat flux type. J Therm Anal Calorim. 2016;123:1795–808.
55. Schawe JEK. Temperature correction at high heating rates for conventional and fast differential scanning calorimetry. Thermochim Acta. 2021;698:178879.
56. Brandts JF, Erickson L, Lysko K, Schwartz AT, Taverna RD. Calorimetric studies of the structural transitions of the human erythrocyte membrane. The involvement of spectrin in the A transition. Biochemistry. 1977;16:3450–4.
57. D'Alessandro A, Anastasiadi AT, Tzounakas VL, Nemkov T, Reisz JA, Kriebardis AG, et al. Red Blood Cell Metabolism In Vivo and In Vitro. Metabolites. 2023;13:793.
58. Bordbar A, Yurkovich JT, Paglia G, Rolfsson O, Sigurjónsson ÓE, Palsson BO. Elucidating dynamic metabolic physiology through network integration of quantitative time-course metabolomics. Sci Rep. 2017;7:46249.
59. Nemkov T, Sun K, Reisz JA, Yoshida T, Dunham A, Wen EY, et al. Metabolism of Citrate and Other Carboxylic Acids in Erythrocytes As a Function of Oxygen Saturation and Refrigerated Storage. Front Med [Internet]. 2017 [cited 2025 Apr 29];4. Available from: https://www.frontiersin.orghttps://www.frontiersin.org/journals/medicine/articles/10.3389/fmed.2017.00175/full
60. Hjellvik V, Sakshaug S, Strøm H. Body mass index, triglycerides, glucose, and blood pressure as predictors of type 2 diabetes in a middle-aged Norwegian cohort of men and women. Clin Epidemiol. 2012;4:213–24.
61. Makhoul Z, Kristal AR, Gulati R, Luick B, Bersamin A, O'Brien D, et al. Associations of obesity with triglycerides and C-reactive protein are attenuated in adults with high red blood cell eicosapentaenoic and docosahexaenoic acids. Eur J Clin Nutr. 2011;65:808–17.
62. Lischka J, Schanzer A, Baumgartner M, de Gier C, Greber-Platzer S, Zeyda M. Tryptophan Metabolism Is Associated with BMI and Adipose Tissue Mass and Linked to Metabolic Disease in Pediatric Obesity. Nutrients. 2022;14:286.
63. Dumont LJ, AuBuchon JP, Collaborative BE for ST (BEST). Evaluation of proposed FDA criteria for the evaluation of radiolabeled red cell recovery trials. Transfusion (Paris). 2008;48:1053–60.
64. D'Alessandro A, Zimring JC, Busch M. Chronological storage age and metabolic age of stored red blood cells: are they the same? Transfusion (Paris). 2019;59:1620–3.
65. Roubinian NH, Reese SE, Qiao H, Plimier C, Fang F, Page GP, et al. Donor genetic and nongenetic factors affecting red blood cell transfusion effectiveness. JCI Insight [Internet]. 2022 [cited 2025 Mar 26];7. Available from: https://insight.jci.org/articles/view/152598
66. Nemkov T, Stefanoni D, Bordbar A, Issaian A, Palsson BO, Dumont LJ, et al. Blood donor exposome and impact of common drugs on red blood cell metabolism. JCI Insight. 6:e146175.
67. Vamvakas EC, Blajchman MA. Transfusion-related immunomodulation (TRIM): An update. Blood Rev. 2007;21:327–48.
68. Silliman CC, Fung YL, Ball JB, Khan SY. Transfusion-related acute lung injury (TRALI): Current Concepts and Misconceptions. Blood Rev. 2009;23:245–55.
69. La Carpia F, Wojczyk BS, Annavajhala MK, Rebbaa A, Culp-Hill R, D'Alessandro A, et al. Transfusional iron overload and intravenous iron infusions modify the mouse gut microbiota similarly to dietary iron. Npj Biofilms Microbiomes. 2019;5:1–11.
70. Cazzola R, Rondanelli M, Russo-Volpe S, Ferrari E, Cestaro B. Decreased membrane fluidity and altered susceptibility to peroxidation and lipid composition in overweight and obese female erythrocytes. J Lipid Res. 2004;45:1846–51.
71. Nan L, Jiang Z, Wei X. Emerging microfluidic devices for cell lysis: a review. Lab Chip. 2014;14:1060–73.
72. Lyu J, Gu Z, Zhang Y, Vu HS, Lechauve C, Cai F, et al. A glutamine metabolic switch supports erythropoiesis. Science. 2024;386:eadh9215.
73. Whillier S, Garcia B, Chapman BE, Kuchel PW, Raftos JE. Glutamine and α-ketoglutarate as glutamate sources for glutathione synthesis in human erythrocytes. FEBS J. 2011;278:3152–63.
74. Wenninger J, Meinitzer A, Holasek S, Schnedl WJ, Zelzer S, Mangge H, et al. Associations between tryptophan and iron metabolism observed in individuals with and without iron deficiency. Sci Rep. 2019;9:14548.
75. Arto C, Rusu EC, Clavero-Mestres H, Barrientos-Riosalido A, Bertran L, Mahmoudian R, et al. Metabolic profiling of tryptophan pathways: Implications for obesity and metabolic dysfunction-associated steatotic liver disease. Eur J Clin Invest. 2024;54:e14279.

76. Bartlett AL, Romick-Rosendale L, Nelson A, Abdullah S, Luebbering N, Bartlett J, et al. Tryptophan metabolism is dysregulated in individuals with Fanconi anemia. Blood Adv. 2021;5:250–61.
77. Cussotto S, Delgado I, Anesi A, Dexpert S, Aubert A, Beau C, et al. Tryptophan Metabolic Pathways Are Altered in Obesity and Are Associated With Systemic Inflammation. Front Immunol [Internet]. 2020 [cited 2025 Apr 29];11. Available from: https://www.frontiersin.orghttps://www.frontiersin.org/journals/immunology/articles/10.3389/fimmu.2020.00557/full
78. Kumar Sarker S, Islam MT, Sarower Bhuyan G, Sultana N, Begum MstN, Al Mahmud-Un-Nabi M, et al. Impaired acylcarnitine profile in transfusion-dependent beta-thalassemia major patients in Bangladesh. J Adv Res. 2018;12:55–66.
79. Zhang Z, Tai Y, Liu Z, Pu Y, An L, Li X, et al. Effects of d-ribose on human erythrocytes: Non-enzymatic glycation of hemoglobin, eryptosis, oxidative stress and energy metabolism. Blood Cells Mol Dis. 2023;99:102725.
80. Yamaguchi T, Fukuzaki S. ATP effects on response of human erythrocyte membrane to high pressure. Biophys Physicobiology. 2019;16:158–66.
81. Binns HC, Alipour E, Sherlock CE, Nahid DS, Whitesides JF, Cox AO, et al. Amino acid supplementation confers protection to red blood cells before Plasmodium falciparum bystander stress. Blood Adv. 2024;8:2552–64.
82. Otoyama I, Hamada H, Kimura T, Namba H, Sekikawa K, Kamikawa N, et al. L-cysteine improves blood fluidity impaired by acetaldehyde: In vitro evaluation. PLOS ONE. 2019;14:e0214585.
83. Raftos JE, Whillier S, Kuchel PW. Glutathione Synthesis and Turnover in the Human Erythrocyte: ALIGNMENT OF A MODEL BASED ON DETAILED ENZYME KINETICS WITH EXPERIMENTAL DATA. J Biol Chem. 2010;285:23557–67.
84. Parra M, Stahl S, Hellmann H. Vitamin B6 and Its Role in Cell Metabolism and Physiology. Cells. 2018;7:84.
85. Reisz JA, Earley EJ, Nemkov T, Key A, Stephenson D, Keele GR, et al. Arginine metabolism is a biomarker of red blood cell and human aging. Aging Cell. 2024;e14388.
86. Roy MK, Cendali F, Ooyama G, Gamboni F, Morton H, D'Alessandro A. Red Blood Cell Metabolism in Pyruvate Kinase Deficient Patients. Front Physiol [Internet]. 2021 [cited 2025 Apr 29];12. Available from: https://www.frontiersin.orghttps://www.frontiersin.org/journals/physiology/articles/10.3389/fphys.2021.735543/full
87. Key AM, Earley EJ, Tzounakas VL, Anastasiadi AT, Nemkov T, Stephenson D, et al. Red blood cell urate levels are linked to hemolysis in vitro and post-transfusion as a function of donor sex, population and genetic polymorphisms in SLC2A9 and ABCG2. Transfusion (Paris). 2025;65:560–74.
88. De Bandt JP, Coumoul X, Barouki R. Branched-Chain Amino Acids and Insulin Resistance, from Protein Supply to Diet-Induced Obesity. Nutrients. 2023;15:68.
89. Rodriguez CJ, Daviglus ML, Swett K, González HM, Gallo LC, Wassertheil-Smoller S, et al. Dyslipidemia Patterns among Hispanics/Latinos of Diverse Background in the United States. Am J Med. 2014;127:1186-1194.e1.
90. Sumner AE, Cowie CC. Ethnic differences in the ability of triglyceride levels to identify insulin resistance. Atherosclerosis. 2008;196:696–703.
91. D'Alessandro A, Fu X, Kanias T, Reisz JA, Culp-Hill R, Guo Y, et al. Donor sex, age and ethnicity impact stored red blood cell antioxidant metabolism through mechanisms in part explained by glucose 6-phosphate dehydrogenase levels and activity. Haematologica. 2021;106:1290–302.
92. Prieto ML, Oralkan Ö, Khuri-Yakub BT, Maduke MC. Dynamic Response of Model Lipid Membranes to Ultrasonic Radiation Force. Phillips W, editor. PLoS ONE. 2013;8:e77115.
93. Blanco-González A, Marrink SJ, Piñeiro Á, García-Fandiño R. Molecular insights into the effects of focused ultrasound mechanotherapy on lipid bilayers: Unlocking the keys to design effective treatments. J Colloid Interface Sci. 2023;650:1201–10.
94. Phillips R, Ursell T, Wiggins P, Sens P. Emerging roles for lipids in shaping membrane-protein function. Nature. 2009;459:379–85.
95. Cahalan SM, Lukacs V, Ranade SS, Chien S, Bandell M, Patapoutian A. Piezo1 links mechanical forces to red blood cell volume. Nathans J, editor. eLife. 2015;4:e07370.